\documentclass[prb,twocolumn,superscriptaddress]{revtex4}

\usepackage{amsmath}
\usepackage{amsfonts}
\usepackage{amssymb,mathrsfs}
\usepackage{graphicx}

\newcommand{\bs}[1]{\boldsymbol{#1}}

\begin{document}

\title{Hydrodynamic approach to electronic transport in graphene}

\author{B.N. Narozhny}

\affiliation{Institute for Theoretical Condensed Matter Physics,
  Karlsruhe Institute of Technology, 76128 Karlsruhe, Germany}
\affiliation{National Research Nuclear University MEPhI (Moscow
  Engineering Physics Institute), Kashirskoe shosse 31, 115409 Moscow,
  Russia}

\author{I.V. Gornyi}

\affiliation{Institute for Theoretical Condensed Matter Physics,
  Karlsruhe Institute of Technology, 76128 Karlsruhe, Germany}
\affiliation{Institute of Nanotechnology, Karlsruhe Institute of
  Technology, 76021 Karlsruhe, Germany}
\affiliation{Ioffe Physical Technical Institute, 194021 St. Petersburg, Russia}

\author{A.D. Mirlin}

\affiliation{Institute for Theoretical Condensed Matter Physics,
  Karlsruhe Institute of Technology, 76128 Karlsruhe, Germany}
\affiliation{Institute of Nanotechnology, Karlsruhe Institute of
  Technology, 76021 Karlsruhe, Germany}
\affiliation{Petersburg Nuclear Physics Institute, 188300 St. Petersburg, Russia}

\author{J. Schmalian}

\affiliation{Institute for Theoretical Condensed Matter Physics,
  Karlsruhe Institute of Technology, 76128 Karlsruhe, Germany}
\affiliation{Institute for Solid State Physics, Karlsruhe Institute of
  Technology, 76021 Karlsruhe, Germany}

\begin{abstract}
 The last few years have seen an explosion of interest in hydrodynamic
 effects in interacting electron systems in ultra-pure materials. In
 this paper we briefly review the recent advances, both theoretical
 and experimental, in the hydrodynamic approach to electronic
 transport in graphene, focusing on viscous phenomena, Coulomb drag,
 non-local transport measurements, and possibilities for observing
 nonlinear effects.
\end{abstract}

\maketitle


\noindent
Hydrodynamics describes a great variety of phenomena around (and
inside) us, including, e.g., the flow of water in rivers, seas, and
oceans, atmospheric phenomena, aircraft motion, the flow of petroleum
in pipelines, or the blood flow through blood vessels in humans and
animals. It has been realized a long time ago that the flow of
electrons in a conductor should, under certain circumstances, also
obey the laws of hydrodynamics. In particular, Gurzhi
\cite{Gurzhi,GurzhiUFN} predicted that electrons can exhibit a
Poiseuille-type flow \cite{Poiseuille,dau6} analogous to that of
liquids in pipes. This should result in an initial power-law decrease
of resistivity with the transverse cross-section of a sample and
temperature, leading to a pronounced minimum. It has turned out,
however, that an experimental realization of such a regime in a metal
or a semiconductor is a highly non-trivial task. Three decades have
passed before de Jong and Molenkamp \cite{Molenkamp} observed the
Gurzhi effect, and even in that work the magnitude of the effect did
not exceed 20\%.

Why is it so difficult to implement electron hydrodynamics in a
laboratory experiment? In contrast to molecules of a conventional
liquid, electrons move in the environment formed by the crystal
lattice. Therefore, the electrons experience not only collisions among
themselves, but also scatter off thermally excited lattice vibrations
-- phonons -- as well as various lattice imperfections (impurities).
The hydrodynamic regime is realized when the frequency of
electron-electron collisions is much larger than the rates of both,
electron-phonon and electron-impurity scattering. These two
requirements limit the temperature window for the hydrodynamic flow
both from above and from below, and may even be in a conflict with
each other. In a typical solid, the elastic impurity scattering
dominates electronic transport at low temperatures, whereas at high
temperatures the leading mechanism is the electron-phonon
scattering. Thus, the requirement of the electron-electron scattering
being the fastest process -- which is the key condition for the
hydrodynamics -- may only be satisfied, if at all, in an intermediate
temperature range. It turns out that this regime is not well developed
in conventional conductors, with the possible exception of the
ultrahigh-mobility {\it GaAs} quantum wells
\cite{Haug,Wegscheider,Zudov} exhibiting negative magnetoresistance
\cite{Alekseev} and ultra-pure palladium cobaltate \cite{Mackenzie}.

The experimental discovery of graphene \cite{Geim} has given a new
boost to the research in the field of quantum transport. In
particular, it has been realized that, among other remarkable
properties, graphene is an excellent material for the realization of
hydrodynamic flow of electrons \cite{Bandurin,Crossno,Geim17}. The
reasons for this are as follows. First, electron-phonon scattering in
graphene on suitable substrates is rather weak, which manifests itself
in unprecedentedly high carrier mobilities at elevated
temperatures. Second, the progress in fabrication allows to produce
graphene samples of outstanding quality (i.e. with a very low
concentration of impurity atoms or lattice defects)
\cite{pon,Decker}. An additional twist to the story is provided by the
``quasi-relativistic'' excitation spectrum of graphene near charge
neutrality -- that of massless 2D Dirac fermions. A fluid emerging in
this situation is formed by two species of carriers with opposite
charges -- electrons and holes \cite{Foster}.

The emergent two-fluid hydrodynamics in graphene is neither Galilean-,
nor Lorentz-invariant
\cite{Muller2,Muller3,Fritz,Schmalian,Tomadin,Svintsov}. The former
feature is a consequence of the linearity of the excitation spectrum
yielding the strong dependence of the transport scattering time in
graphene on electron-electron interaction \cite{Schutt}. The latter
appears due to the classical (and moreover, three-dimensional) nature
of the Coulomb interaction between charged quasiparticles in graphene.
As a result, rather than assuming the standard approach
\cite{Tomadin,Svintsov,dau10} the hydrodynamic equations have to be
derived from the microscopic theory
\cite{Hartnoll,Muller2,Muller3,us1,us2}.

Microscopically, the motion of the charge carriers may be described
using the methods of the kinetic theory \cite{dau10}. At the simplest
level, transport properties of conventional metals and semiconductors
are determined by means of a perturbative solution of the kinetic
equation under the assumption of a weak external bias (i.e. the
electric field or temperature gradient). Final results are typically
expressed in terms of the linear relations between the macroscopic
currents and the external fields, e.g. Ohm's law. The
proportionality coefficients, such as the electrical resistivity, are
determined by the rates of momentum-nonconserving scattering processes
-- disorder scattering at low temperatures and electron-phonon
scattering at high temperatures. Inelastic scattering processes
responsible for the equilibration of the system are assumed to be much
faster that any transport-related time scale. In contrast, the
hydrodynamic behavior appears when the dominant scattering process in
the system is due to electron-electron interaction. In conventional
systems, this may happen either in pure Fermi liquids \cite{Khalat},
clean samples in constricted geometries \cite{Gurzhi}, or in
low-density 2D electron systems at high enough
temperatures \cite{Spivak}.

In graphene, the solution of the kinetic equation is simplified by the
kinematic peculiarity of electron-electron scattering leading to a
formal divergence of the collision integral known as the ``collinear
scattering singularity''
\cite{Muller2,Muller3,us1,us3,Kashuba,Schutt,Brida}. This is another
consequence of the Dirac-like spectrum: the energy and momentum
conservation laws are identical for quasiparticles moving in the same
direction. Although the divergence is regularized by dynamical
screening \cite{us1,Schutt,Brida}, the remaining separation in the
values of the scattering rates allows for a nonperturbative solution
to the kinetic equation facilitating the transition from the
microscopic theory to the hydrodynamic description \cite{us2}.

The two-fluid hydrodynamic approach can be extended to double-layer
systems \cite{us3} used in Coulomb drag measurements
\cite{Narozhny,Gorbachev,Tutuc,Tutuc2,Titov}. The resulting theory
yields a quantitative description of giant magnetodrag in graphene at
charge neutrality \cite{Titov}. Practical calculations within this
approach involve solving the hydrodynamic equations taking into account
boundary conditions at the sample edges uncovering the key role of
quasiparticle recombination processes in transport measurements in
mesoscopic-sized samples. Qualitatively, one can interpret the results
of this theory in terms of a semiclassical two-band model
\cite{us4}. Treating this model phenomenologically, one can extend the
hydrodynamic theory to further 2D two-component material near charge
neutrality. In particular, the theory suggests an explanation for the
phenomenon of linear magnetoresistance in nearly compensated materials
in classically strong magnetic fields \cite{us5}.

The interest in hydrodynamic behavior of charge carriers in graphene
is fueled by a promise for potential applications
\cite{Maier,Grigorenko}, as well as the exciting prospect of observing
phenomena otherwise belonging to the realms of high-energy and plasma
physics in a condensed matter laboratory \cite{Levitov13}.  Recent
progress in experimental techniques
\cite{Basov,Koppens,Herrero,Fogler,Gonzalez,Koppens15} brings the
studies of nonlinear phenomena \cite{us2} within reach.  Nonlocal
transport phenomena are already a subject of intensive research
\cite{Abanin,Bandurin,Levitov,GeimTheory}.  At the same time, the
connection between the nearly relativistic hydrodynamics in graphene
\cite{Hartnoll} and the holography approach to quantum field theory
\cite{Sachdev11,Sachdev2016} represents a promising avenue for
theoretical developments.

In this paper, we briefly review the current status of the fast
developing field of electron hydrodynamics in graphene. We begin with
the overview of the theory, focusing on the symmetry properties of
Dirac quasiparticles in graphene and consequent peculiarities of the
derivation of the hydrodynamic equations. Then we discuss several
examples of applications of the hydrodynamic approach to
experimentally relevant problems including linear magnetoresistance
and giant magnetodrag. Finally, we discuss the recent experiments on
hydrodynamic flow of graphene. We conclude by discussing the possible
extension of the hydrodynamic theory onto further solid state systems
with nearly relativistic spectra, e.g. Weyl and Dirac semimetals.

\section{Electronic transport in graphene: from microscopic theory to hydrodynamics}

Traditional hydrodynamics \cite{dau6} describes the system in terms of
the velocity field $\bs{v}$. The equations describing the velocity
field (e.g., the Euler equation in the case of the ideal liquid or the
Navier-Stokes equation if dissipation is taken into account) can be
either inferred from symmetry arguments or derived from the Boltzmann
kinetic equation. Both approaches require one to express the fluxes of
conserved quantities (energy, momentum, etc.) in terms of $\bs{v}$.
In particular, the viscous terms appearing in the Navier-Stokes
equation can be traced to a particular approximation for the momentum
flux (or the stress tensor) $\Pi_{\alpha\beta}$. The specific form of
$\Pi_{\alpha\beta}$ depends on whether one discusses a usual,
Galilean-invariant or a relativistic, Lorentz-invariant system. As
graphene possesses neither symmetry, the precise form of the
hydrodynamic equations has to be derived from the microscopic kinetic
theory.

\subsection{Kinetic equation}

Microscopically, the electronic system can be described by the Boltzmann
kinetic equation
\begin{equation}
\mathcal{L}f = St_{ee}[f] - \tau_\text{dis}^{-1}(f-\langle f\rangle_\varphi),
\label{BE}
\end{equation}
with the standard Liouvillian form in the left-hand side,
\begin{equation}
\mathcal{L} = \partial_t + \bs{v}\cdot\bs{\nabla}_{\bs{r}} 
+ \left[e\bs{E}+e(\bs{v}\times\bs{B})\right]\cdot\bs{\nabla}_{\bs{k}}, 
\label{Liouvillian}
\end{equation}
and the collision integral in the right-hand side. Impurity scattering
can be described within the usual $\tau$-approximation with
$\tau_\text{dis}$ being the disorder mean free time. The collision
integral $St_{ee}[f]$ describes electron-electron interaction. Note
that in Eq.~(\ref{Liouvillian}) we consider only the orbital effect of
the magnetic field. The Zeeman splitting is small and plays no role in
the context of quasiclassical transport in relatively weak magnetic
fields.

The strength of the electron-electron interaction is typically
described by the effective interaction constant,
${\alpha_g=e^2/\epsilon{v}_g}$, where $\epsilon$ is the effective
dielectric constant of the substrate and $v_g$ is the quasiparticle
velocity in graphene, ${v_g=|\bs{v}|}$. Depending on the substrate,
the coupling constant may be small \cite{koz,Sheehy,Titov},
${\alpha_g<1}$. In that case, the solution of the kinetic equation
(\ref{BE}) is facilitated by the so-called collinear scattering
singularity. Indeed, scattering of quasiparticles with the collinear
velocities results in a divergence
\cite{Muller2,Muller3,us1,us3,Kashuba,Schutt,Brida} in
$St_{ee}[f]$. The formal divergence is regularized by dynamical
screening \cite{us1,Schutt,Brida} leaving generic relaxation rates
finite, but large, ${\tau_g^{-1}\propto|\ln\alpha_g|\gg1}$. At the
same time, macroscopic currents related to conserved quantities
(i.e. the energy current $\bs{j}_E$ and the electric current
$\bs{j}$), as well as the so-called imbalance current \cite{Foster}
$\bs{j}_I$, are relaxed at much longer time scales. This scale
separation is the key point allowing for a nonperturbative solution of
the kinetic equation (\ref{BE}).

In graphene, the energy current $\bs{j}_E$ is proportional to the
momentum and hence cannot be relaxed by electron-electron interaction.
Therefore, the steady state cannot be established without some
extrinsic mechanism (except for the neutrality point, where in the
absence of magnetic field the steady state can be established by
Coulomb interaction alone). Therefore, weak disorder scattering has to
be taken into account. The relative weakness of potential disorder 
in the hydrodynamic regime is characterized by the condition
\begin{equation}
  \label{iddr}
\tau_{ee}\ll\tau_\text{dis}.
\end{equation}
Similarly, any other scattering process, such as electron-phonon,
quasiparticle recombination, or three-particle processes, should also
be characterized by time scales that are much longer than $\tau_{ee}$.

Scale separation in $St_{ee}[f]$ results in the two-step
thermalization in graphene. Short-time scattering processes
(characterized by $\tau_g$) establish the so-called ``unidirectional
thermalization'' \cite{us3}: the collinear scattering singularity
implies that the electron-electron interaction is more effective along
the same direction. At this point one can solve the kinetic equation
within linear response \cite{us1}. The resulting non-equilibrium
distribution function is proportional to the three macroscopic
currents, $\bs{j}$, $\bs{j}_E$, and $\bs{j}_I$, which can be found
from the macroscopic equations obtained by integrating
Eq.~(\ref{BE}). This approach is reviewed in the next subsection.

At longer time scales, scattering processes characterized by
$\tau_{ee}$ thermalize quasiparticles with different directions of
their velocity and hence establish the local equilibrium \cite{us2}.
Consequently, at longer times the system can exhibit a true
hydrodynamic behavior described by nonlinear equations analogous to
the classical Navier-Stokes equation. Linearizing these equations one
recovers the macroscopic equations of the linear response theory.

\subsection{Macroscopic linear response theory}

Electronic transport in graphene within linear response can be
described in terms of the three macroscopic currents in the system
\cite{us1}: the energy current $\bs{j}_E$, the electric current
$\bs{j}$, and the imbalance current (or total quasiparticle flow)
$\bs{j}_I$. While physically inequivalent, the three 2D vectors cannot
be linearly independent. It is not surprising, that several two-fluid
descriptions of graphene considered either $\bs{j}$ and $\bs{j}_E$
\cite{Hartnoll,Muller2,Fritz,us3}, or $\bs{j}$ and $\bs{j}_I$
\cite{Foster,Titov}. In particular, in the so-called Fermi-liquid
regime of very high doping all three currents are equivalent. In this
simplest case, the theory is reduced to the standard macroscopic
electrodynamics (ultimately, Ohm's law) that can be derived
perturbatively \cite{us6}.

The theory also simplifies at charge neutrality, where relatively
simple equations exhibit all qualitative features of the
theory. Consider first an infinite system (meaning that the sample
size is much larger than any dynamic length scale in the system). Then
the quasiparticle system is essentially uniform and can be described
by the following equations (which essentially generalize the Ohm's law
to the case of neutral graphene):
\begin{subequations}
\label{sleq3dp}
\begin{equation}
\label{sljeq3dp}
\bs{E} + {\cal R}_H\bs{K}\times \bs{e}_{\bs{B}}
= {\cal R}_0 \bs{j} + \frac{\pi\bs{j}}{2e^2 T\tau_{vv}\ln 2},
\end{equation}
\begin{eqnarray}
\label{slqeq3dp}
{\cal R}_H\bs{j}\times \bs{e}_{\bs{B}}
= {\cal R}_0 \frac{e}{2T\ln 2}\bs{j}_E,
\end{eqnarray}
\begin{eqnarray}
\label{slpeq3dp}
{\cal R}_H\bs{j}\times \bs{e}_{\bs{B}}
= e{\cal R}_0 \bs{j}_I + \frac{\pi\bs{C}}{2e^2 T\tau_{ss}\ln 2}.
\end{eqnarray}
\end{subequations}
Here $\bs{e}_{\bs{B}}$ is the unit vector in the direction of the
magnetic field, the coefficients with dimensions of resistivity are
\[
  {\cal R}_0 = \frac{\pi}{2\ln2 e^2T\tau_\text{dis}}, \quad
  {\cal R}_H = \frac{\pi\omega_B}{2\ln2 e^2T}, \quad
  \omega_B = \frac{ev_g^2B}{2\ln2 cT},
\]
the vector $\bs{C}$ is a linear combination
\[
\bs{C} = e\bs{j}_E \frac{\gamma_0}{2T\Delta(0)\ln 2}
- e\bs{j}_I\frac{{\cal N}_2(0)}{\Delta(0)}, 
\]
with the numerical coefficients
\[
\gamma_0 = \frac{\pi^2}{12\ln^22} \approx 1.7119, 
\quad
{\cal N}_2(0) = \frac{9\zeta(3)}{8\ln^32}\approx 4.0607,
\]
and
\[
\Delta(0) = \gamma_0^2 - {\cal N}_2(0) \approx -1.1303,
\]
the scattering rates $\tau_{vv}^{-1}$ and $\tau_{ss}^{-1}$ describe
the mutual scattering in the velocity (electric current) and imbalance
channels (in the Fermi liquid limit the rate $\tau_{vv}^{-1}$ and the
vector $\bs{C}$ vanish due to restored Galilean invariance; as a
result, electron-electron interaction has no effect on electrical
resistivity), and finally the Lorentz term in Eq.~(\ref{sljeq3dp}) is
determined by the vector
\[
\bs{K} = \bs{j}\times\bs{e}_{\bs{B}}\frac{\kappa{\cal R}_H}{{\cal R}_0\Delta(0)},
\]
where
\[
\kappa = \gamma_0-1
+ \left[\gamma_0-{\cal N}_2(0)\right]
\frac{\Delta(0) -  \gamma_0\tau_\text{dis}/\tau_{ss}}
{\Delta(0)-{\cal N}_2(0)\tau_\text{dis}/\tau_{ss}}<0.
\]

With the above vector $\bs{K}$, the direction of the Lorentz term in
Eq.~(\ref{sljeq3dp}) coincides with the direction of $\bs{j}$. Thus,
no classical Hall voltage can be induced at the Dirac point (as
expected from symmetry considerations)
\begin{equation}
\label{res-rhdp}
R_{H}(\mu=0) = 0.
\end{equation}

The equations (\ref{sleq3dp}) describe positive magnetoresistance in
neutral graphene: the longitudinal magnetoresistance exhibits a
quadratic $B$-dependence (here both $\kappa$ and $\Delta(0)$ are
negative, so their ratio is positive) \cite{Muller2,us1}
\begin{eqnarray}
&&
R(B;\mu=0) = R(B=0;\mu=0) + \delta R(B;\mu=0),
\nonumber\\
&&
\nonumber\\
&&
\delta R(B;\mu=0) = \frac{{\cal R}_H^2\kappa}{{\cal R}_0\Delta(0)}
\propto \frac{v_g^4\tau_\text{dis}}{c^2} \frac{B^2}{T^3}.
\label{res-rbdp}
\end{eqnarray}
Physically, this result follows form the fact that at charge
neutrality electron-electron scattering involves carriers from both
bands in graphene. In this sense, the positive magnetoresistance
(\ref{res-rbdp}) is similar to the well-known classical
magnetoresistance in multiband semiconductors.

If the sample is doped away from neutrality, the role of the second
band gradually diminishes until the single-band degenerate limit (the
Fermi liquid limit) is reached. At this point the classical Hall
effect is restored, while the magnetoresistance (\ref{sleq3dp})
vanishes. Given the inverse proportionality between the classical Hall
coefficient and the carrier density, ${R_H(\mu\gg T)\sim 1/(nec)}$,
the Hall coefficient exhibits a maximum at ${\mu\sim T}$ in agreement
with experiment \cite{Novoselov}.

Finally, the equation (\ref{slqeq3dp}) becomes meaningless in the
absence of disorder scattering, i.e. for ${\cal R}_0=0$. This is a
manifestation of the fact that the electron-electron interaction is
insufficient to establish the steady state in the system: without
extrinsic scattering mechanisms, the energy current (proportional to
the total momentum conserved in electron-electron scattering
processes) will increase indefinitely under external bias.

\subsection{Nonlinear hydrodynamic equations for charge carriers in graphene}
\label{nlh}

The linear response theory reviewed in the previous subsection is
valid at all time scales greater than $\tau_g$. At the same time, in
the interaction-dominated regime (\ref{iddr}) there is a wide window
of parameters where the quasiparticle flows in graphene are truly
hydrodynamic.

Indeed, at time scales of order $\tau_{ee}\gg\tau_g$ the
electron-electron scattering processes lead to thermalization of
quasiparticles moving in different direction yielding a transition
from unidirectional thermalized state to local equilibrium. The latter
is described by the distribution function \cite{us2,Svintsov}
\begin{equation}
f^{(0)}_{\lambda,\bs{k}}(\bs{r})
\!=\! \left\{1\!+\!
\exp\left[\beta(\bs{r})(\epsilon_{\lambda,\bs{k}}\!-\!\mu_\lambda(\bs{r})
  \!-\!\bs{u}(\bs{r})\!\cdot\!\bs{k})\right]
\right\}^{-1}\!,
\label{loc_eq}
\end{equation}
where ${\epsilon_{\lambda,\bs{k}}=\lambda v_g k}$ denotes the energies
of the electronic states with the momentum $\bs{k}$ in the band
${\lambda=\pm1}$, ${\mu_{\lambda}(\bs{r})}$ the local chemical
potential, the local temperature is encoded in
${\beta(\bs{r})=1/T(\bs{r})}$, and ${\bs{u}(\bs{r})}$ is the
hydrodynamic velocity. The latter should not be confused with the
quasiparticle velocity $\bs{v}$ and is defined by its relation to the
conserved hydrodynamic current in the system, i.e. the energy current
[this relation can be found by substituting the distribution function
  (\ref{loc_eq}) into the microscopic definition of the energy
  current; hereafter we use the units with ${v_g=1}$]
\begin{equation}
\bs{j}_E =\frac{3n_E\bs{u}}{2+u^2},
\label{jE}
\end{equation}
where ${u=|\bs{u}|}$.

The particular form of Eq.~(\ref{loc_eq}) reflects the symmetry
properties of the two-particle (electron-electron) scattering:
conservation of energy, momentum, and the particle number in each band
independently. The latter is strictly speaking violated by Auger
processes, quasiparticle recombination, and three-particle collisions,
but all such processes are assumed to be weak, i.e. characterized by
very long scattering times (at least of order $\tau_{\rm
  dis}$). Neglecting these processes for the time being, one arrives
at the three continuity equations for the hydrodynamic densities
\begin{subequations}
\label{cont_eq}
\begin{equation}
\partial_t n + \bs{\nabla}\cdot\bs{j} = 0,
\label{cont_n} 
\end{equation}
\begin{equation}
\partial_t n_I + \bs{\nabla}\cdot\bs{j}_I = 0,
\label{cont_nI} 
\end{equation}
\begin{equation}
\partial_t n_E + \bs{\nabla}\cdot\bs{j}_E = e\bs{E}\cdot\bs{j},
\label{cont_nE} 
\end{equation}
\end{subequations}
as well as the equation for the energy current
\begin{equation}
\partial_t j_{E,\alpha}+\nabla_\beta\Pi^E_{\beta\alpha}
-e n E_\alpha-en(\bs{u}\times\bs{B})_\alpha
= -j_{E,\alpha}/\tau_\text{dis}.
\label{eom_jE}
\end{equation}
The equation (\ref{eom_jE}) generalizes the usual starting point for
derivation of hydrodynamic equations: one has to relate the momentum
flux $\Pi^E_{\alpha\beta}$ to the hydrodynamic velocity $\bs{u}$. In
graphene this can be done with the help of Eq.~(\ref{loc_eq}):
\begin{equation}
\Pi^E_{\alpha\beta}=\frac{n_E}{2+u^2}
\left[\delta_{\alpha\beta}(1-u^2)+3u_\alpha u_\beta\right]
+\delta\Pi^E_{\alpha\beta}.
\label{PiE}
\end{equation}
The last term $\delta\Pi^E$ describes the dissipative effects not
included in Eq.~(\ref{loc_eq}). The first term is the generalization
of the standard stress tensor of an ideal liquid \cite{dau6} to the
case of Dirac fermions. Its unusual form reflects the absence of
Galilean and Lorentz symmetries in graphene.

Furthermore, in contrast to the standard hydrodynamics of an
electrically neutral fluid \cite{dau6} the continuity equation for the
energy density (\ref{cont_nE}) contains the ``source term'' physically
corresponding to Joule's heat. This term (naturally absent in the
linear response theory) connects the different macroscopic currents
reflecting the multiband nature of graphene. In terms of the
hydrodynamic velocity, the electric and imbalance currents are given by
\begin{subequations}
\label{j}
\begin{equation}
\bs{j} = n\bs{u} + \delta\bs{j},
\label{EOS_j} 
\end{equation}
\begin{equation}
\bs{j}_I = n_I\bs{u} + \delta\bs{j}_I,
\label{EOS_jI}
\end{equation}
\end{subequations}
where again the dissipative corrections $\delta\bs{j}$,
$\delta\bs{j}_I$ are introduced.

Finally, the electric field in Eqs.~(\ref{cont_nE}) and (\ref{eom_jE})
should include the self-consistent Vlasov field
\begin{equation}
\bs{E}_{V}(\bs{r})= -\bs{\nabla}_r\!\int\!d^2r' V(\bs{r}-\bs{r}') \delta n(\bs{r}'),
\label{Vlasov}
\end{equation}
where ${\delta n(\bs{r})=n(\bs{r})-n_0}$ is the fluctuation of the
local charge density, $n_0$ is the background charge density, and
${V(\bs{r})=e^2/r}$ is the (3D) Coulomb potential.

Neglecting the dissipative corrections $\delta\bs{j}$,
$\delta\bs{j}_I$, and $\delta\Pi^E_{\alpha\beta}$, the equations
presented in this subsection describe the ideal flow of electronic
fluid in graphene \cite{us2} generalizing the usual Euler equation
\cite{dau6}. Taking into account dissipative processes one arrives at
the generalized Navier-Stokes equation \cite{us2}.

The dissipative corrections $\delta\bs{j}$ and $\delta\bs{j}_I$ can be
found to leading order in the gradient expansion \cite{us2} and
have the form
\begin{subequations}
  \label{diss_corr}
\begin{equation}
\begin{pmatrix}
\delta\bs{j} \\
\delta\bs{j}_I
\end{pmatrix}
=
{\mathcal{C}}^{-1}_J\bs{\nu}_J,
\end{equation}
where the vector $\bs{\nu}_J$ is given by
\begin{equation}
\bs{\nu}_J\!=\!
\begin{pmatrix}
\frac{n}{3n_E}\bs{\nabla} n_E \!-\! \frac{1}{2}\bs{\nabla} n \!-\! 
\left[\frac{2en^2}{3n_E}\!-\!\frac{e}{2}\partial_\mu n\right]\!\bs{E} \\
\frac{n_I}{3n_E}\bs{\nabla} n_E \!-\! \frac{1}{2}\bs{\nabla} n_I \!-\! 
\left[\frac{2enn_I}{3n_E}\!-\!\frac{e}{2}\partial_\mu n_I\right]\!\bs{E}
\end{pmatrix},
\label{nu_matrix}
\end{equation}
and the matrix ${\mathcal{C}}_J$ is the reduced collision integral
(within the same three-mode approximation as in the previous
subsection). Its inverse is given by
\begin{equation}
{\mathcal{C}}^{-1}_J=
\begin{pmatrix}
\tau_{1} & \tau_{2} \\
\tau_{3} & \tau_{4}
\end{pmatrix},
\label{coll_int_matrix}
\end{equation}
\end{subequations}
where $\tau_j$ are the transport scattering times. The off-diagonal
times $\tau_{2,3}$ change their sign under ${n\rightarrow-n}$. In the
non-degenerate regime ${\mu\ll T}$ the times $\tau_j$ are determined
by temperature and electron-electron interaction,
${\tau_j={f_j(\mu/T)}/(\alpha_g^2T)}$, where ${f_j(\mu/T)}$ is a
smooth, dimensionless function. Close to the neutrality point,
\begin{subequations}
\label{taui}
\begin{equation}
\tau_{2}=\tau_{3}=0,
\end{equation} 
while
\begin{equation}
\tau_{1}^{-1}
\approx 2.22 \: \alpha_g^2 T,
\end{equation}
and 
\begin{equation}
\tau_{4}^{-1} 
\approx 0.05 \: \alpha_g^2 T.
\end{equation}
\end{subequations}
In the degenerate regime, ${\mu\gg{T}}$, the system behaves similarly
to the usual Fermi liquid, where all macroscopic currents are
equivalent and electron-electron interaction does not affect transport
Physically, this happens due to the restored Galilean invariance,
while technically the interaction-induced transport relaxation rate
${\sim{T}^4/\mu^3}$ (which is the same for all currents) is much
smaller than the rate ${\tau_{ee}^{-1}\sim{T}^2/\mu}$ determining the
quasiparticle lifetime as well as thermalization.

To leading order in weak deviations from local equilibrium
(${\delta n_E/n_E\ll1}$, ${T\delta n_I/n_E\ll1}$, and ${T\delta
  n/n_E\ll1}$), the correction $\delta\Pi^E$ takes the canonical form
\cite{dau6}
\begin{equation}
\delta\Pi^E_{\alpha\beta}=
-\eta \left[\nabla_{\alpha} u_\beta + \nabla_{\beta} u_\alpha 
- \delta_{\alpha\beta}\bs{\nabla}\!\cdot\!\bs{u}\right] ,
\label{dPI_E}
\end{equation}
with $\eta$ being the viscosity coefficient (from the viewpoint of the
hydrodynamic theory this relation is the definition of viscosity; note
the absence of the bulk viscosity in this theory). Close to the
neutrality point \cite{us2}, the viscosity is given by 
\begin{equation}
  \label{vis}
  \eta = T(\tau_{\pi,1} n + \tau_{\pi,2} n_I)/4 + 3 \tau_{\pi,3}n_E/8,
\end{equation}
where the scattering times $\tau_{\pi,i}$ are obtained from the
collision integral \cite{us2} similarly to the above times $\tau_i$,
see Eq.~(\ref{coll_int_matrix}). At charge neutrality, the first term
in Eq.~(\ref{vis}) does not contribute ($\tau_{\pi,1}=0$), while the
remaining time scales $\tau_{\pi,2(3)}\sim1/(\alpha_g^2T)$ are of the
same order as Eqs.~(\ref{taui}). As a result \cite{Schmalian,us2},
\begin{equation}
  \eta(\mu=0)\sim T^2/\alpha_g^2.
\end{equation}
In the degenerate regime, $\mu\gg T$, the usual Fermi-liquid viscosity
is recovered \cite{dau6,prin,us2}
\begin{equation}
  \eta(\mu\gg T) \propto 1/T^2.
\end{equation}

The expressions for the dissipative correction in terms of the
hydrodynamic velocity and densities allow one to close the set of the
hydrodynamic equations. In particular, the generalized Navier-Stokes
equation can be written in the canonical form \cite{us2,Schmalian}
\begin{eqnarray}
\label{NS}
&& 
\!\!\!\!W\partial_t\bs{u} \!+\! W(\bs{u}\cdot\nabla)\bs{u}\!+\!\nabla P \!+\! \bs{u}\partial_t P 
\!+\! \bs{u}(\delta\bs{j}\cdot\bs{E}) 
\\
&&
\nonumber\\
&&
\qquad\qquad\qquad\qquad\qquad\quad   
= \!
en[\bs{E}\!-\!\bs{u}(\bs{u}\cdot\bs{E})] \!+ \!\eta\nabla^2\bs{u},
\nonumber
\end{eqnarray}
where the hydrodynamic pressure is
\begin{equation}
P = \frac{(1-u^2)n_E}{2+u^2},
\label{pressure}
\end{equation}
and the enthalpy of the system ${W=n_E+P}$ is given by
\begin{equation}
W = \frac{2w}{2+u^2},\quad
w=n_E+P_0=3n_E/2,
\label{enthalpie}
\end{equation}
with the latter being the linear enthalpy of graphene. For small
velocities, the pressure assumes the standard value for a scale
invariant gas, ${P_0=n_E/2}$. For large velocities ${u\lesssim1}$ it
vanishes as ${\sim(1-u^2)}$.

The generalized Navier-Stokes equation (\ref{NS}), the continuity
equations (\ref{cont_eq}), the Vlasov self-consistency (\ref{Vlasov}),
and the expressions for the dissipative corrections (\ref{dPI_E}) and
(\ref{diss_corr}) constitute the complete hydrodynamic description of
charge carriers in graphene.

\section{Signatures of hydrodynamic behavior in graphene}

\subsection{Longitudinal conductivity in monolayer graphene}
\label{sigma}

As discussed above, a highly nontrivial feature of graphene is that
inelastic electron-electron collisions may limit the conductivity at
the Dirac point without any disorder or phonon
scattering~\cite{Kashuba,Muller2,Fritz,Foster,Schutt}.  This
peculiarity of graphene -- which should be contrasted to conventional
systems where interactions do not lead (in the absence of Umklapp
scattering) to finite resistivity -- is a consequence of the
particle-hole symmetry and decoupling between velocity and momentum.
As a result, although the total momentum of interacting particles is
conserved during inelastic collisions, the total current may
relax. The collision-limited conductivity of undoped graphene is found
\cite{Kashuba,Muller3,Schutt,us2} to be inversely proportional to
$\alpha_g^2$
\begin{equation}
  \label{sdp}
  \sigma(\mu=0)\approx\frac{e^2}{h}\frac{0.76}{\alpha_g^2},
\end{equation}
and depends on temperature only through the renormalization of
$\alpha_g$ \cite{Sheehy}. Potential disorder introduces relaxation of
the quasiparticle velocity affecting the transport scattering rate due
to electron-electron scattering. Moreover, in the presence of disorder
the density of states at the Dirac point becomes nonzero. As a result,
the conductivity (\ref{sdp}) is modified to
\begin{equation}
  \label{sdpd}
  \sigma(\mu=0)\sim \frac{e^2}{h}\frac{T+\tau_\text{dis}^{-1}}{\alpha_g^2T+\tau_\text{dis}^{-1}},
\end{equation}
Away from the Dirac point, the dc conductivity of graphene diverges in
the absence of extrinsic scattering, since the momentum and velocity
modes are no longer orthogonal to each other.  Still, however,
inelastic electron-electron collisions that establish the hydrodynamic
regime play an important role for transport in clean graphene samples,
in particular, in suspended graphene.

\begin{figure}
\centerline{
  \includegraphics[width=0.9\columnwidth]{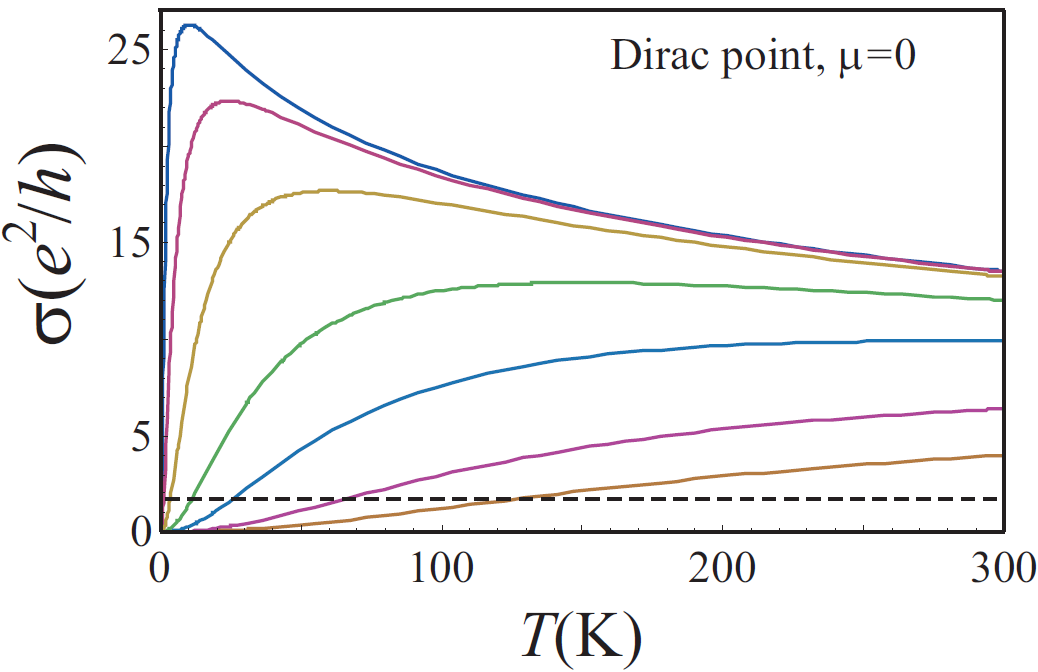}
  }
 \caption{Temperature dependence of conductivity of suspended graphene
   at the Dirac point; impurity concentration $n_i$ grows from top to
   bottom. Adapted from Ref.~\onlinecite{Kachor}.}\label{Fig:7}
\end{figure}

Transport properties of suspended graphene \cite{Bolotin} were
explored in Ref.~\onlinecite{Kachor}, with a particular focus on the
case of zero chemical potential.  The interplay of electron-electron
collisions with the disorder-induced scattering and interaction with
flexural (out-of-plane deformation) phonons was analyzed.  The
temperature dependence of the conductivity, Fig. \ref{Fig:7}, is
governed by the electron-impurity (lowest $T$), electron-electron
(intermediate $T$), and electron-phonon (highest $T$) scattering. The
hydrodynamic description is thus valid in a rather broad range of
intermediate temperatures.

\subsection{Thermal transport in graphene}

In a recent paper, Crossno and co-workers \cite{Crossno} have studied
the electric and thermal transport in high-quality monolayer graphene
samples encapsulated in hexagonal boron nitride. They have evaluated
the Lorentz number characterizing the ratio of the thermal
conductivity to the electric one. In conventional transport regime,
the Lorentz number takes a universal value depending on fundamental
constants only. This reflects the fact that the relaxation of charge
current and of the thermal current is provided by the same scattering
processes. Indeed, Crossno et al. do find this universal value in the
most part of the parameter plane spanned by the temperature and the
charge carrier density. However, in a relatively narrow range of
concentrations around zero (i.e., for the chemical potential being
near the Dirac point of particle-hole symmetry) and at intermediate
temperatures between 40 and 100 K, the Lorentz number is found to be
much larger than the universal value, with the enhancement factor
reaching 22. This remarkable observation is a clear manifestation of
the hydrodynamic behavior of charge carriers in graphene, namely, of
formation of the electron-hole ``Dirac fluid''.

An intuitive explanation of the strong enhancement of Lorentz number
in the Dirac fluid regime is as follows. When an electric field is
applied, electrons and holes move in opposite directions, and the
friction between them (mediated by electron-electron interaction)
leads to a finite electric resistivity, see Sec.~\ref{sigma}. On the
other hand, when a temperature gradient is applied, the electrons and
holes move in the same direction, implying no friction and thus
infinite thermal conductivity. An accurate theoretical analysis
\cite{Fritz,Foster,Schmalian,Principi} confirms this conclusion.

The fact that the hydrodynamic behavior emerges only in an
intermediate temperature range (between 40 and 100 K in the experiment
of Ref.~\onlinecite{Crossno}) is in full agreement with the discussion in
the introductory part of the paper. At low temperatures the transport
is dominated by impurity scattering, while at high temperatures the
phonon scattering becomes the main scattering mechanism.

Another signature of the hydrodynamic behavior in graphene is the
substantial enhancement of a related physical quantity, namely the
thermoelectric power \cite{Hartnoll,Kim} in high-mobility graphene
devices. In contrast to the earlier experiments using graphene on
silicon oxide \cite{Kim2009}, the new measurements showed that at
relatively high temperatures the thermoelectric power significantly
exceeds the standard Mott relation \cite{Mott} approaching the ideal
hydrodynamic limit\cite{Fritz,Foster,Muller3,Xie}, where (in the
absence of disorder) the thermopower equals the thermodynamic entropy
per carrier charge. This limit, however, was not reached in the
experiment \cite{Kim}, presumably due to inelastic scattering of
charge carriers off the optical phonons which appears to be
non-negligible even at temperatures much lower than the phonon
frequency. The latter effect demonstrates the upper limit of the
temperature window for the hydrodynamic behavior.

Finally, one might expect a similar enhancement of the thermal
transport in graphene subjected to an external magnetic field. In the
earlier experiments \cite{Kim2009,Lau,Ong}, the Nernst signal in doped
graphene appeared to be in agreement with the generalized Mott
relation \cite{Girvin,Jonson} and exhibited strong oscillations
\cite{Varlamov}. However, at charge neutrality the measurements
\cite{Kim2009,Ong} show a large enhancement of the Nernst signal
strongly deviating from the Mott relation. The thermoelectric power in
magnetic field exhibited similar behavior. While the observed effects
may be attributed to either the peculiarities of the zeroth Landau
level in graphene \cite{Abanin2007} or the sample shape dependency of
the two-terminal transport measurements near the neutrality point
\cite{Williams}, it would be extremely interesting to study the effect
of the magnetic field on the hydrodynamic thermal transport in
graphene.

\subsection{Viscosity and nonlocal transport}

Currently, manifestations of viscosity in transport properties of
graphene in the hydrodynamic regime attract a great deal of attention.
In a recent experiment, Bandurin et al. \cite{Bandurin} have measured
a negative four-terminal resistance of graphene samples. Specifically,
they observed a negative voltage drop in the vicinity of
current-injection contacts in an intermediate temperature range
(roughly between 100 and 200 K). This remarkable observation provides
an unambiguous evidence in favor of ``whirlpool'' current patterns
(vorticity) in graphene as expected for a viscous fluid
\cite{Levitov,Torre,GeimTheory}.

In a further recent work, Moll et al. \cite{Mackenzie} have observed
an evidence of the viscous transport predicted in \cite{Gurzhi} by
measuring the resistance of restricted channels of the layered (i.e.,
essentially two-dimensional) material, PdCoO$_2$. This material is
known for its purity \cite{Hicks} as evidenced by the extremely small
in-plane resistivity and the large mean-free path. Moreover, the ratio
of the mean free paths inferred from the analysis of the de Haas-van
Alphen effect (a measure of a wide range of scattering processes) and
from the resistivity (a measure of momentum relaxation) is about an
order of magnitude smaller than that in ordinary metals, indicating
that PdCoO$_2$ is a good candidate for a study of hydrodynamic
effects.

\subsection{Linear magnetoresistance}
\label{lmr}

\begin{figure}
\centerline{
\includegraphics[width=0.9\columnwidth]{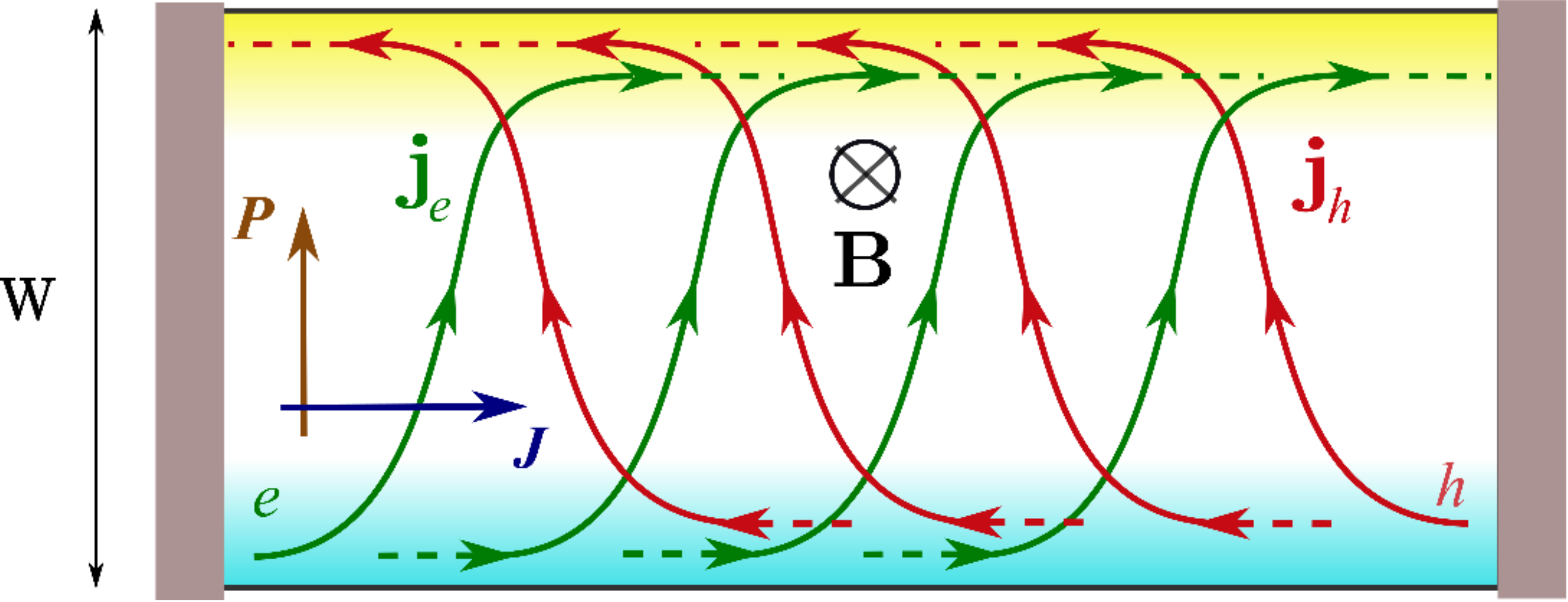}
}
\caption{Electron (green) and hole (red) currents in a finite graphene
  sample. The quasiparticle flow (denoted by $\bs{P}$) results in
  excess quasiparticle density near the sample edges, where
  recombination processes due to electron-phonon interaction lead to
  linear magnetoresistance. Adapted from Ref.~\onlinecite{us4}.}
  \label{fig:picture}
\end{figure}

\begin{figure}
\centerline{
\includegraphics[width=0.9\columnwidth]{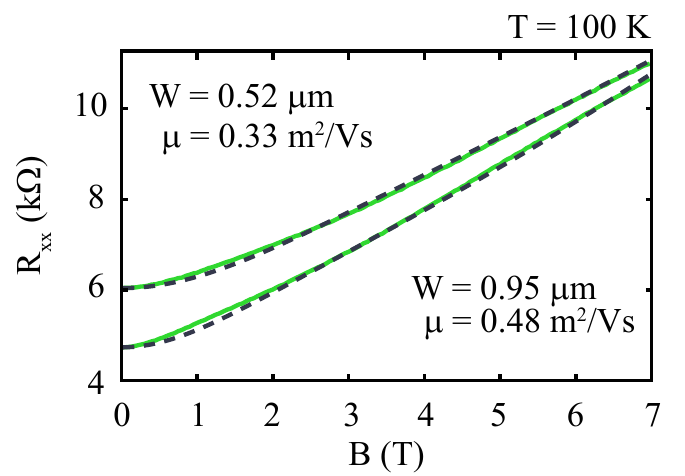}
}
\caption{Magnetoresistance of bilayer graphene measured in two samples
  \cite{us5} with different widths and mobilities at the charge
  neutrality point and $T=100$K. Solid (green) lines: experimental
  data; Dashed (blue) lines: theoretical fit using the semiclassical
  description of Ref.~\onlinecite{us4}. Adapted from
  Ref.~\onlinecite{us5}.}
  \label{fig-haug}
\end{figure}

The hydrodynamic approach to the magnetotransport in infinite graphene
samples predicts a positive parabolic magnetoresistance,
Eq.~(\ref{res-rbdp}). However, many magnetotransport experiments in
multilayer graphenes exhibit linear magnetoresistance in sufficiently
strong magnetic fields \cite{Friedman,Liao,Weber}. In
Refs.~\onlinecite{us4,us1,us5,us17} it was demonstrated that boundary
effects may lead to a non-saturating classical linear
magnetoresistance at the charge neutrality point when the sample width
is comparable with the electron-hole recombination length $\ell_0$.

The mechanism of this boundary effect in the magnetoresistance in a
two-component fluid is illustrated in Fig.~\ref{fig:picture} for an
electron-hole symmetric system at charge neutrality, where the Hall
effect for electrons is compensated by that for holes. The
distribution of electron and hole currents, $\bs{j}_{e,h}$, differs
substantially in the bulk of the sample and near the edges.

The main contribution to the magnetoresistance originates in the
narrow edge regions with the width of the order of the recombination
length $\ell_{R}$, where both the electron and hole currents are
directed essentially along the edge. The edge contribution to the
overall resistance at charge neutrality is proportional to
$L/\ell_{R}$, where $L$ is the sample length. With increasing magnetic
field, the recombination length gets shorter because of multiple
cyclotron returns of electron and holes to each other:
$\ell_{R}\propto 1/|B|$.  The total sheet resistance $R_\square$ of
the sample with the width $W\gg\ell_{R}$ can be be estimated by
treating the edge and the bulk as parallel resistors:
$R_\square^{-1}=(R^{-1}_\textrm{bulk}+R^{-1}_\textrm{edge})L/W$.  For
sufficiently strong magnetic field $B$, the magnetoresistance is then
linear in $B$:
\begin{equation}
\label{estimate}
R_\square=\frac{1}{e\rho}\frac{W}{\ell_{0}} |B|,
\end{equation}
where $\rho$ is the total quasiparticle density.  Away from the charge
neutrality, a finite Hall effect leads to a saturation of the linear
magnetoresistance. This qualitative derivation \cite{us4} is in
agreement with the rigorous calculation for graphene \cite{us1,us17}
based on the three-mode hydrodynamic model.  This mechanism of linear
magnetoresistance was measured in recent experiments on bilayer
graphene \cite{us5}, see Fig.~\ref{fig-haug}.

\subsection{Coulomb drag in graphene}

Hydrodynamic ideas can also be applied to double-layer systems
\cite{Narozhny,Titov,Apostolov,Chen}. Consider a sample consisting of
two graphene sheets (or layers) separated by an insulator, that is
thick enough, so that the two layers are electrically isolated (i.e.,
there is essentially no electron tunneling between them), but at the
same time not too thick, so that the electric field created by
electrons in one layer penetrates into the other. In this case, the
interlayer Coulomb interaction leads to an observable effect known as
the Coulomb drag \cite{Narozhny}: if an electric current, $I_1$, is
passed through one of the layers, then a current -- or a voltage,
$V_2$, in the case of an open circuit -- will be generated in the
other layer. The measured drag coefficient (the ratio of the induced
voltage to the driving current, ${R_D=-V_2/I_1}$) is extremely
sensitive to the microscopic structure of the electronic system in the
two layers, making it an important tool for experimental studies of
many body systems.

\begin{figure}
\centerline{
\includegraphics[width=0.85\columnwidth]{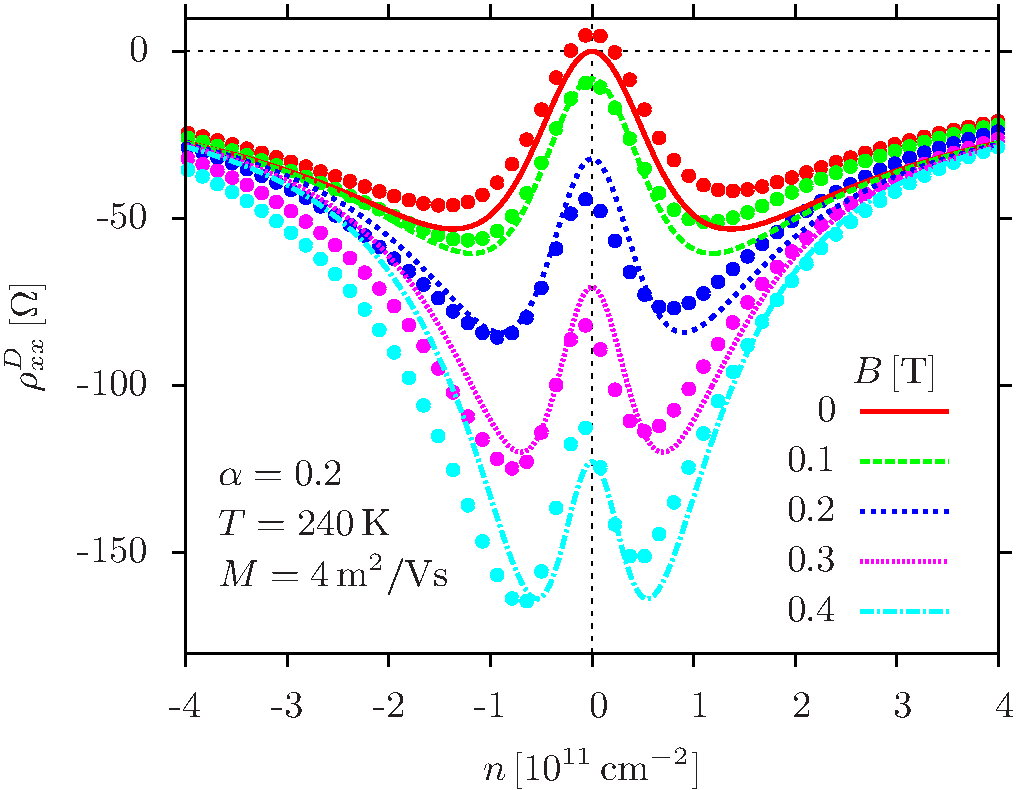}
}
\caption{Giant magnetodrag in graphene. Solid symbols represent the
  experimental data measured at $T=240$K. The curves represent the
  results of theoretical calculations. Both graphene sheets are kept
  at the same carrier density. Adapted from Ref.~\onlinecite{Titov}.}
  \label{fig-mdr}
\end{figure}

\begin{figure}[t]
\centering
  \includegraphics[width=0.85\columnwidth]{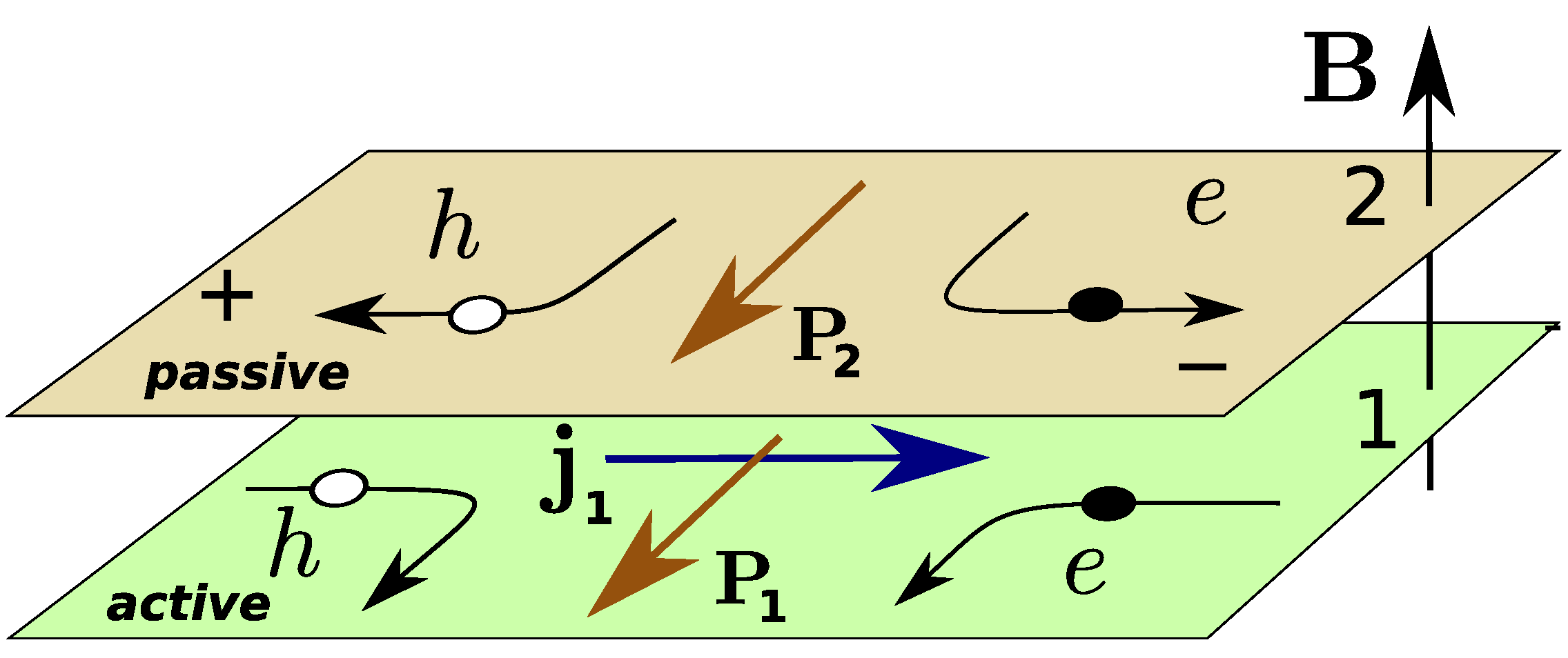}\\
  \includegraphics[width=0.85\columnwidth]{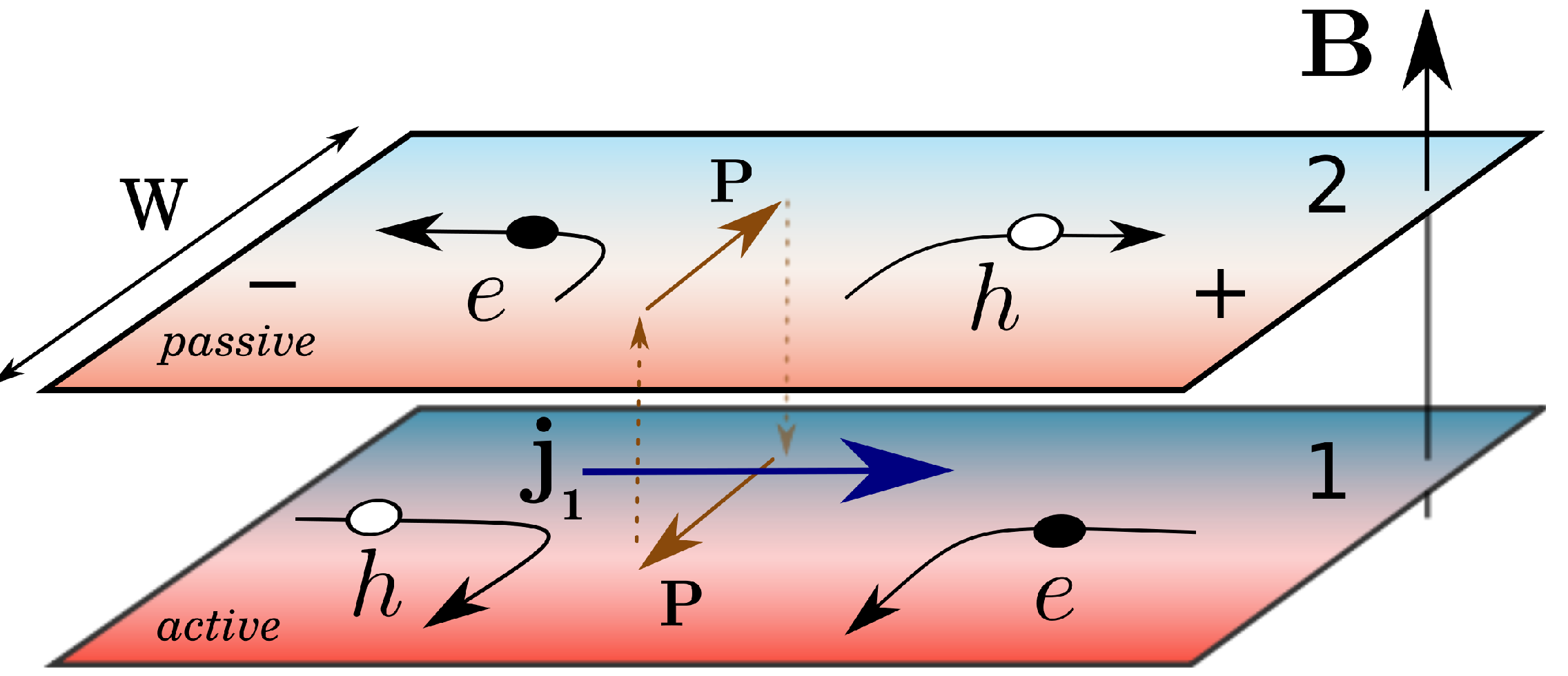}
\caption{Qualitative mechanism of magnetodrag in graphene. Top panel:
  infinite sample, where the lateral quasiparticle flow (denoted by
  $\bs{P}_i$) is transferred between the layers by the Coulomb
  interaction. Bottom panel: finite-size sample, where the total
  quasiparticle flow $\bs{P}_1+\bs{P}_2$ in the sample vanishes due to
  hard wall boundary conditions; this leads to the quasiparticle flows
  in the two layers having opposite directions, $\bs{P}_1=-\bs{P}_2$,
  which leads to negative magnetodrag. Adapted from
  Ref.~\onlinecite{Titov}.}
  \label{fig-md}
\end{figure}

The main physical properties distinguishing graphene from conventional
conductors -- the absence of Galilean invariance, collinear scattering
singularity, precise electron-hole symmetry at charge neutrality,
unidirectional thermalization, and most importantly, the two-band
carrier system -- have their direct manifestation in Coulomb drag.  In
ultra-clean graphene close to charge neutrality, the hydrodynamic
description yields non-trivial, temperature-independent drag resistivity 
\begin{equation}
  \rho_D \sim \frac{\alpha_g^2}{e^2}\frac{\mu_1\mu_2}{\mu_1^2+\mu_2^2},
\end{equation}
which remains finite if the chemical potentials of the two layers,
$\mu_1$ and $\mu_2$, approach zero simultaneously. However, this
result does not survive if even infinitesimal disorder is present in
the system, in which case the leading-order contribution to drag
vanishes due to the electron-hole symmetry.

The inequivalence of the electric current and macroscopic
quasiparticle flow in graphene leads to the effect of giant
magnetodrag \cite{Titov}, see Fig.~\ref{fig-mdr}. The effect can be
traced back to the fact that the Lorentz force in the electron and
hole bands has the opposite sign, which is also the reason for the
anomalously large Nernst effect \cite{Hartnoll,Kim2009,Lau,Ong} in
monolayer graphene and vanishing Hall effect at charge
neutrality. Magnetodrag can be qualitatively understood by considering
the two-band transport similar to that discussed in Sec.~\ref{lmr}
above, see Fig.~\ref{fig-md}. The driving current in a drag experiment
corresponds to the counter-propagating flow of electrons and holes,
which precisely at the neutrality point is characterized by zero total
momentum. This is the physical reason for vanishing drag in the
presence of electron-hole symmetry. However, if a weak magnetic field
is applied, electrons and holes are deflected by the Lorentz force
yielding the neutral quasiparticle flow lateral to the driving
current. Now this flow does carry a nonzero momentum which can be
transferred to the second layer by the interlayer interaction. There,
the Lorentz forces acting on the two types of carriers will drive the
charge flow parallel to the driving current and hence yield nonzero
magnetodrag. The sign of the effect depends on the geometry of the
system.

\subsection{Nonlinear hydrodynamic phenomena}

Nonlinear phenomena (along with the viscous effects discussed below)
are the hallmark of the hydrodynamic flow. However, nonlinear effects
do not often feature in condensed matter laboratory experiments, where
the vast majority of transport measurements are performed within
linear response. A notable exception is the experiment of
Ref.~\onlinecite{Molenkamp}, where signatures of the Gurzhi effect
were observed in the nonlinear current-voltage characteristic.

\begin{figure}
\centerline{
\includegraphics[width=0.9\columnwidth]{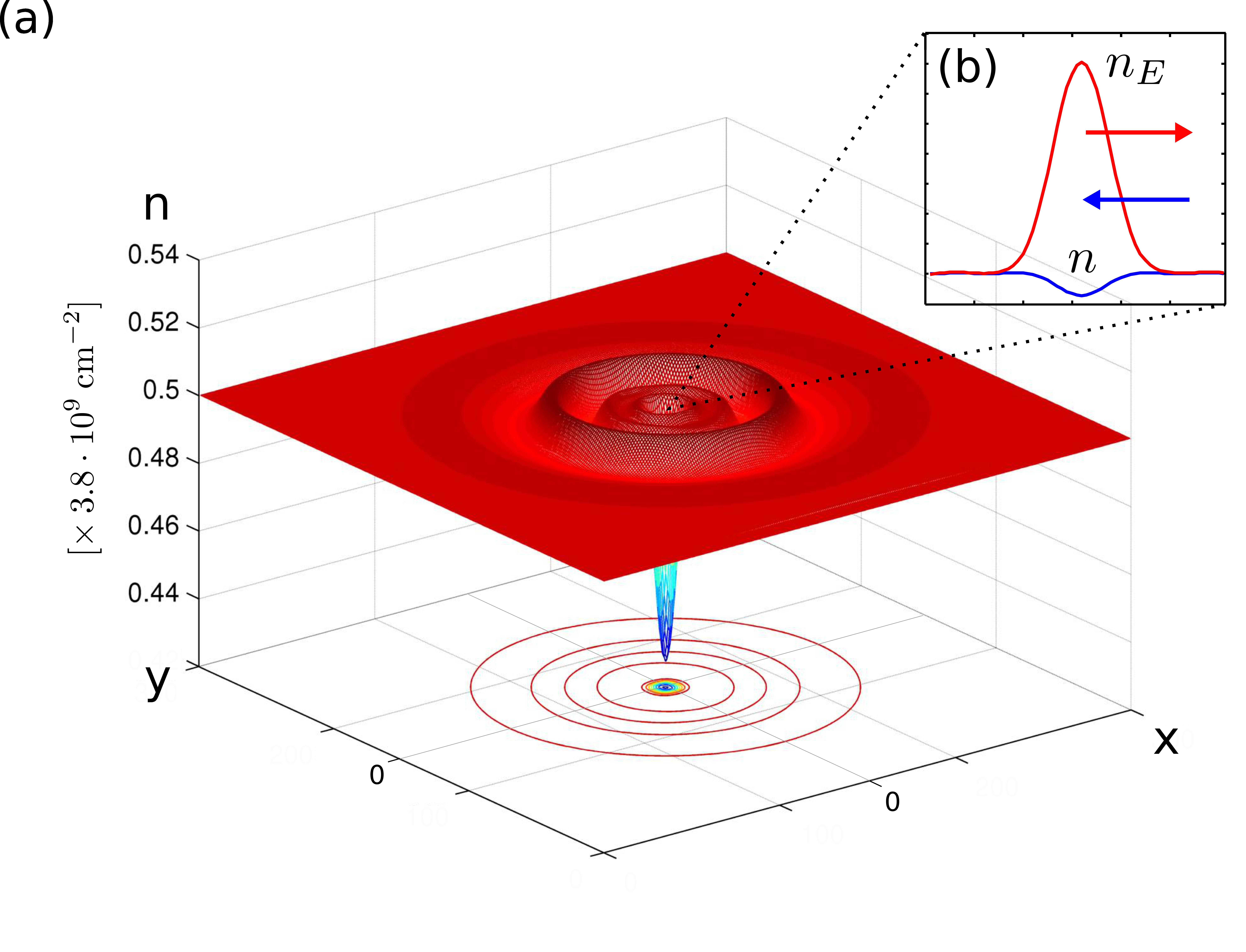}
}
\caption{Hot spot in graphene. Inset: the soliton-like composite
  object at the origin. The blue curve shows the dip in the charge
  density, while the red curve shows the excess energy density. The
  red arrow indicates the pressure force compensated by the
  self-consistent electric field (shown by the blue arrow). Adapted
  from Ref~\onlinecite{us2}.}
  \label{fig-hs}
\end{figure}

\begin{figure}
\centerline{
\includegraphics[width=0.9\columnwidth]{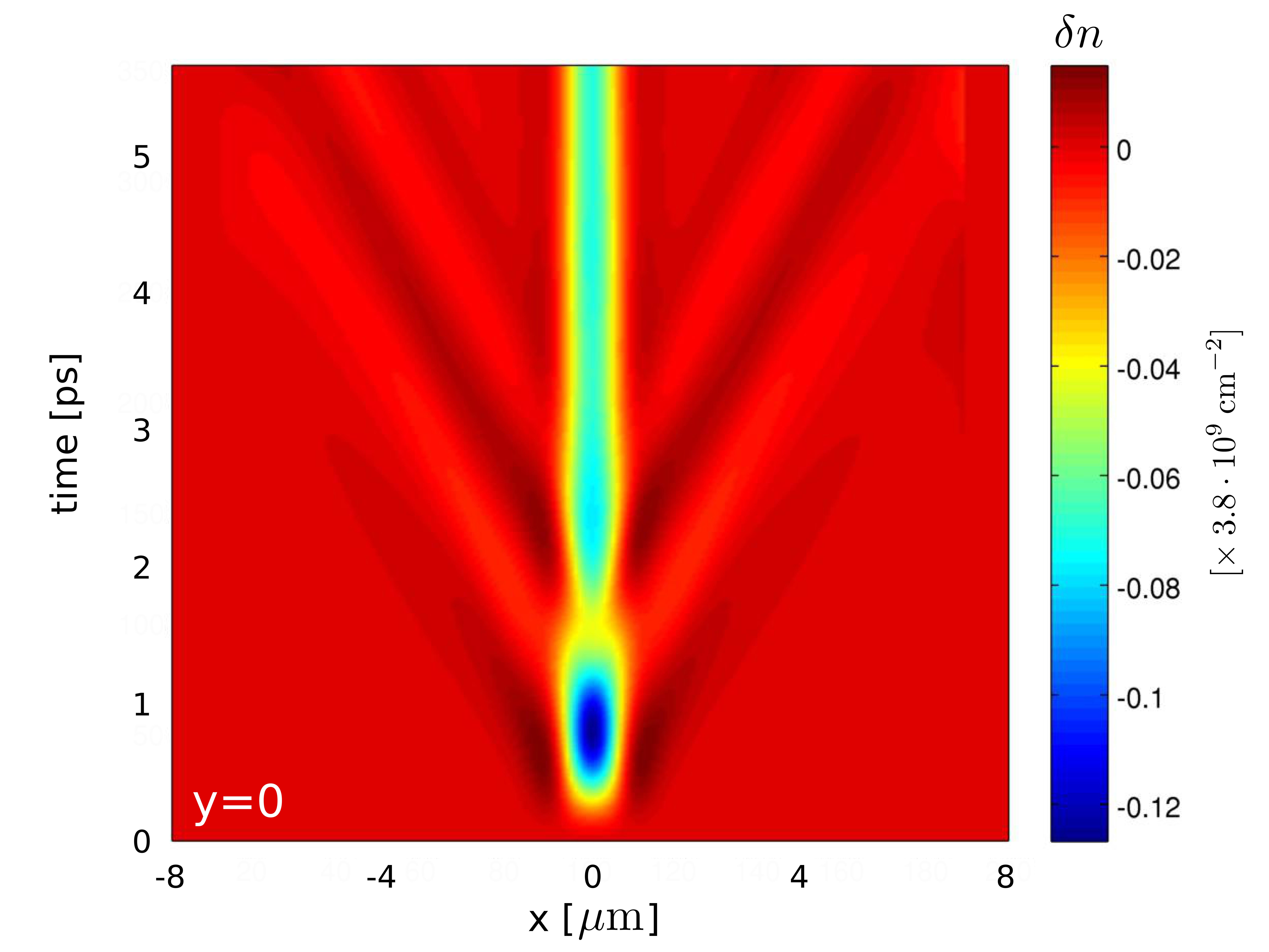}
}
\caption{Hot spot relaxation. The color-map plot shows the time
  evolution of the charge density as a function of $x$ along the line
  $y=0$ for short enough time scales. Adapted from Ref.~\onlinecite{us2}.}
  \label{fig-hst}
\end{figure}

Theoretically, a representative example of nonlinear physics in
graphene -- relaxation of a hot spot -- was discussed in
Ref.~\onlinecite{us2}, see Figs.~\ref{fig-hs} and \ref{fig-hst}. A hot
spot is a nonequilibrium state characterized by a locally elevated
quasiparticle energy density. The hot spot can be created using a
local probe \cite{Basov} or a focused laser beam \cite{Koppens}. The
experiments of Refs.~\onlinecite{Basov,Koppens} were devoted to the
study of plasmon propagation in graphene. In these experiments the
samples were continuously illuminated by the external field. In
contrast, one can use a single pulse to create nonequilibrium energy
density profile and follow the subsequent time evolution. As expected
\cite{Basov,Koppens}, the hot spot immediately emits plasmonic
waves. However, the nonlinear theory of Sec.~\ref{nlh} yields an
additional, rather surprising result: a nonzero excess energy density
remains at the hot spot accompanied by a nonzero excess charge
density.

In graphene, the hydrodynamic energy flow is coupled to the charge
flow by means of Vlasov self-consistency (\ref{Vlasov}). At charge
neutrality, the two flows may decouple and the hydrodynamic equation
(\ref{NS}) admits solutions in the form of energy waves
\cite{Fogler16,us2}. In the absence of dissipation and disorder, the
energy wave is acoustic, similar to the ``cosmic sound'' (i.e. the
long-wavelength oscillations in interacting relativistic models)
\cite{Levitov13}. In a charged system, the self-consistent electric
field ensures that the energy waves are accompanied by fluctuations of
the charge density. Physically, the excess energy density leads to the
appearance of the pressure force,
${\nabla_\alpha\Pi^E_{\alpha\beta}}$, which in turn initiates the
hydrodynamic flow. The corresponding electric current carries charge
away from the hot spot leading to a depletion of the charge density.
The appearing nonequilibrium profile creates the self-consistent
electric field. Remarkably, in the absence of dissipation the
generated electric force compensates the above pressure force. As a
result, one finds a stable, soliton-like solution with a composite --
energy and charge -- density profile at the hot spot, see the inset in
Fig.~\ref{fig-hs} (here the initial perturbation of the energy density
was chosen to have a form of a Gaussian with the peak height nearly
double the unperturbed value, ${n_E=1.8n_E^{(0)}}$ leading to the
micrometer size of the soliton). Dissipation leads to decay of the
quasi-stable soliton, but this decay occurs at time scales that are
long compared to the initial emission of the plasmon waves, see
Fig.~\ref{fig-hst}. The plasmon waves themselves are damped by viscous
effects.

\section{Final remarks}

Hydrodynamic flow of electrons in solids should be observable not only
in graphene, but in any material that is clean enough to satisfy the
condition (\ref{iddr}). In particular, modern semiconductor technology
allows to fabricate ultrahigh-mobility heterostructures
\cite{Haug,Wegscheider,Zudov}, which should make it possible
\cite{Alekseev} to observe hydrodynamic behavior in conventional
conductors \cite{Spivak,Hruska,Levchenko} for the first time since the
original observation of the Gurzhi effect \cite{Molenkamp}. Very
recently, it was suggested that a viscous hydrodynamic flow in
electronic systems might exhibit enhanced, higher-than-ballistic
conduction \cite{Geim17,Falkovich}.

At the same time, hydrodynamic behavior might be observable in a wide
range of novel materials including the 2D metal palladium cobaltate
\cite{Mackenzie}, topological insulators \cite{Fradkin} (where the
conducting surface states may exhibit hydrodynamic behavior), and Weyl
semimetals. The latter systems have attracted considerable attention
since they exhibit a solid state realization of the Adler-Bell-Jackiw
chiral anomaly \cite{Adler,Bell,Nielsen,Zhang}. One of the hallmark
manifestations of the anomaly in Weyl systems is the recently observed
\cite{Zhang} negative magnetoresistance
\cite{Nielsen,Son,SpivakAndreev,Lucas}. It would be very interesting
to observe relativistic Weyl hydrodynamics \cite{Lucas} in these
systems.

While the findings reviewed in this paper are highly encouraging, we
are only at the beginning of a challenging way towards a better
control and understanding of electron fluid mechanics. A further
improvement of sample quality is desirable to obtain broader
hydrodynamic regimes. It remains to be explored what are possible
applications of the emerging ``hydrodynamic electronics''.

\medskip

\begin{acknowledgments}

We would like to thank P.S. Alekseev, U. Briskot, A.P. Dmitriev,
L. Fritz, A.K. Geim, R.V. Gorbachev, R.J. Haug, Y.L. Ivanov,
V.Y. Kachorovskii, M.I. Katsnelson, A.A. Kozikov, A. Levchenko,
J. Link, M. M\"uller, K.S. Novoselov, P.P. Orth, P.M. Ostrovsky,
L.A. Ponomarenko, S. Sachdev, M. Sch\"utt, D.E. Sheehy, A.V. Shytov,
D. Smirnov, M. Titov, T. Tudorovskiy, G.Y. Vasileva, Y.B. Vasilyev,
and the late A.K. Savchenko for collaboration on the problems reviewed
in this paper. This work was supported by the Deutsche
Forschungsgemeinschaft via the Priority Program 1459 ``Graphene''
(I.V.G and A.D.M).

\end{acknowledgments}


\end{document}